# A Data-Driven Approach to Pre-Operative Evaluation of Lung Cancer Patients

Oleksiy Budilovsky[1], Golnaz Alipour[1], Andre Knoesen[1], Lisa Brown[2], Soheil Ghiasi[1]
[abudilovsky@ucdavis.edu]
[1]Department of Electrical and Computer Engineering, UC Davis
[2]Department of Surgery and Comprehensive Cancer Center, UC Davis

*Abstract*—Lung cancer is the number one cause of cancer deaths. Many early stage lung cancer patients have resectable tumors; however, their cardiopulmonary function needs to be properly evaluated before they are deemed operative candidates. Consequently, a subset of such patients is asked to undergo standard pulmonary function tests, such as cardiopulmonary exercise tests (CPET) or stair climbs, to have their pulmonary function evaluated. The standard tests are expensive, labor intensive, and sometimes ineffective due to co-morbidities, such as limited mobility. Recovering patients would benefit greatly from a device that can be worn at home, is simple to use, and is relatively inexpensive. Using advances in information technology, the goal is to design a continuous, inexpensive, mobile and patient-centric mechanism for evaluation of a patient's pulmonary function. A light mobile mask is designed, fitted with $CO_2$, $O_2$, flow volume, and accelerometer sensors and tested on 18 subjects performing 15 minute exercises. The data collected from the device is stored in a cloud service and machine learning algorithms are used to train and predict a user's activity. Several classification techniques are compared – K Nearest Neighbor, Random Forest, Support Vector Machine, Artificial Neural Network, and Naive Bayes. One useful area of interest involves comparing a patient's predicted activity levels, especially using only breath data, to that of a normal person's, using the classification models.

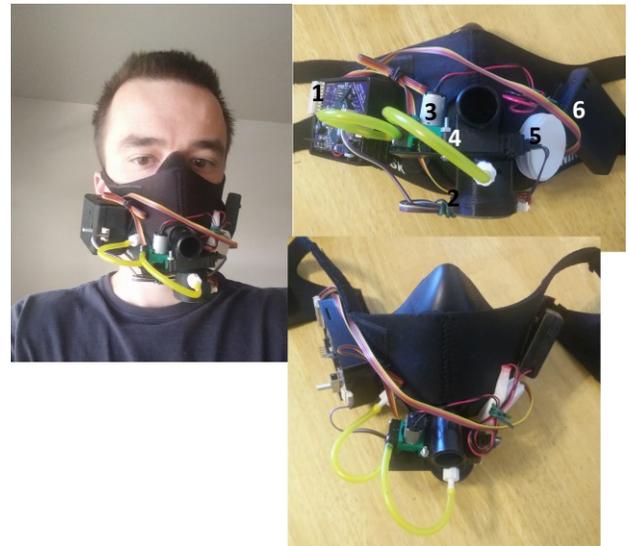

Fig. 1. How the mask looks on the subject, as well as two angles of the mask. Labels are: 1 – PCB in 3D printed enclosure (cap off) with soldered on electronics, 2 – $CO_2$ chamber with sensor at the bottom, 3 – motorized pump for $CO_2$ sampling, 4 – volume flow meter, 5 – oxygen sensor with 3D printed cap, 6 – battery pack for the pump. The yellow tubes carry pumped air from the mask chamber to the motor, and the to the $CO_2$ chamber. An exit hole in the chamber allows the air to escape.

## I. INTRODUCTION

Non-small cell lung cancer (NSCLC) is the number one cause of cancer deaths among both men and women, accounting for approximately 25% of such deaths [1]. Many patients with early stage lung cancer have a resectable tumor, but may not be operative candidates due to comorbidities (obesity, broken leg, etc.) or inadequate pulmonary function. Thus, assessment of the risk of morbidity and mortality prior to resection is important. With proper assessment, appropriate treatment can be selected for the patient.

A Smart Mask is designed to assist with patient evaluation at the patient's convenience, without having to do a hospital visit for a cardiopulmonary exercise test (CPET). Because a CPET is expensive, labor intensive, and sometimes ineffective, a better solution would be to have a light and inexpensive device for the patient's use at home, as well as providing the data for the doctor remotely.

The mask collects $CO_2$, oxygen, respiratory rate and flow, and acceleration, while the patient is instructed to perform various activities. It connects to an Android phone via Bluetooth and saves data collected from the mask to a cloud service. Authorized entities such as doctors can then access the data online.

Section II describes the current state of practice in lung cancer patient evaluation. Section III describes the mask hardware in detail and the purpose of the data analytics. Section IV talks about the specific problem, along with the machine learning algorithms used. Section V is the experimental evaluation section, explaining the details of the data collection, the trials ran, and the data flow paths. This section also lists the results. Finally, section VI concludes the paper and discusses potential areas of interest in the future.

## II. STATE OF PRACTICE IN LUNG CANCER PATIENT EVALUATION

Patients traditionally undergo spirometry and diffusion testing to determine preoperative pulmonary function, which yields $FEV_1$ (forced expiratory volume) and DLCO (diffusing capacity of the lung for CO) measures. These raw values are compared to other individuals of similar age, weight, height and gender to determine the patient's relative health standing.

Both $FEV_1$ and DLCO are independent predictors of morbidity and mortality after lung resection for non-small cell lung cancer [2]. If $FEV_1$ and/or DLCO place the individual in the bottom 40% of the reference population, guidelines recommend that the patient undergo cardiopulmonary exercise testing (CPET) for further evaluations [3].

CPET is a clinically-administered test in which the patient runs on a treadmill while many sensors are attached, such as a facemask, electrodes for ECG, and heart rate sensor [4]. The tubing from the mask connects to a bulky and expensive machine, and then to a gas mixing chamber where $CO_2$ and Oxygen levels are sampled. The patient is asked to gradually increase her physical activity to a level she can sustain. The data collected can be used to stratify patients for risk of morbidity and mortality after lung resection.

Such testing is expensive, labor intensive, and sometimes inaccurate if the patient cannot fully perform the exercises. At this point stair climbing may be used instead of the treadmill; however, stairclimbing carries its own limitation problems due to musculoskeletal or peripheral vascular diseases.

### III. SMART MASK

*A. Project Vision*

Due to recent advances in information technology, a continuous, inexpensive, in-home and patient-centric device can be developed for evaluation of a patient's pulmonary function. A light mobile mask is developed that can be easily taken on and off the face, and collects essential breathing parameters that CPET collects. When the mask is turned on with a switch, a Bluetooth connection is established to an Android smartphone, and data collection can begin. When the trial ends $CO_2$, Oxygen, flow rate and volume, and acceleration data is sent to a cloud service. This "modern day" solution allows patients to perform necessary testing at home and doctors to have near real-time data.

*B. Mask Hardware*

A 2.5" by 1.8" board is designed and printed to house the Programmable System-on-Chips (PSOC) 4 and 5, sensor connections, and remaining circuitry. The PSOC4 allows for connection to a BlueTooth Low Energy Enabled Android phone, sending data in real time. The $O_2$ sensor (SST LOX-O2 model) uses fluorescence quenching to detect changes in oxygen concentration, with a detection range of 0-25%. Because atmospheric levels are around 21% and respiration levels at 16%, this range is sufficient. The MinIR 100% $CO_2$ sensor uses Non-Dispersive Infrared (NDIR) detection, and can measure typical $CO_2$ breath levels (near 4%). Using infrared technology over mass spectrometer is acceptable even though the response time is slower – accuracy can theoretically be achieved up to 0.05% [5]. All the sensors and chips are powered and downregulated with a 9V rechargeable battery. The flow meter (Honeywell AWM730B5) measures pressure difference from either end, and measures up to ±300 liters per minute (close to person's maximal breathing).

The accelerometer reports data in the 3-D coordinate space, and after an internal high pass filter reads zeros while the accelerometer is still.

Both $CO_2$ and $O_2$ sensors allow for polling of data at 2 Hz, whereas the flow and accelerometer have much higher polling capabilities. A sampling rate between 11 Hz and 12 Hz was used for all mask sensors.

Heart rate data is measured with a separate device – a Polar H7 Heart Rate Sensor and FitnessTracker, which straps to the wearer's chest and picks up heart rate at 1 Hz. The sensor pairs with a watch and sends requires syncing post-trial.

The mask is comprised of a rubber face mask and a neoprene cover to hold the rubber in place. The mask contains three holes – the central one holds the flow meter, the right hole contains the oxygen sensor, and the left hole is barricaded with a plastic cap. The $CO_2$ sensor is placed underneath the mask - it proved to saturate during testing when blown on directly, and was placed in a 3D printed chamber and attached to a motorized pump, which moves air from the mask chamber into the smaller $CO_2$ chamber, giving the sensor a constant sampling of exhaled air. A separate battery pack was attached to the right side of the mask to power the pump, powered by three AAA batteries. Specific hardware details for the $CO_2$ and $O_2$ sensors and chip design are omitted here but can be found in [6] (flow meter and sensors rearrangement are newer modifications).

*C. Data Analytics*

Using the data collected from the Smart Mask, one can apply data analytics and machine learning techniques to answer predictive questions. In *machine learning* terms, the concept of prediction stems from the idea of using a set of pre-labeled data derived from a large set of trials to train predictive models. Once the models are sufficiently trained, new data coming from the sensors can be used directly on the models. Such tools provide the opportunity for various hypotheses to test - such as predicting activity using the data, and specifically the breathing data.

### IV. PATIENT ACTIVITY DETECTION USING BREATHING DATA

*A. Problem Statement*

Unlike previous studies that utilize wearable inertial sensors on the body to predict activities [7,8,9], or utilize various sensor networks for health monitoring [10], our primary motivation is to be able to predict a user's activity from the mask data, and more specifically breathing data, from several seconds. If an accurate predictor is regularly predicting that the patient is doing an activity she is not actually doing, that could be an immediate trigger for the monitoring doctor to look more in depth and perhaps call the patient on site for a CPET.

A potential constraint for this mask is not having $CO_2$ and $O_2$ sensor levels be as calibrated as for the in-hospital CPET. Due to the lightweight and mobile nature of the mask, and the fact that there is no mixing chamber as in a CPET, the measurements may not be as accurate as those done with medical hardware.

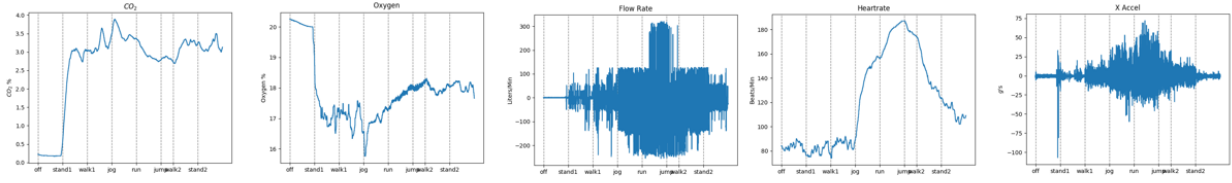

Fig. 3. Randomly selected trial data. *Y* and *z* axes profiles are omitted due to space considerations.

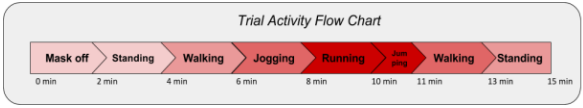

Fig. 2. Order of activities performed, with timestamps.

*B. General Approach*

To train predictive algorithms, first data needs to be collected. We devise an activity order for the patients to perform, and collect trial data. Then the data is split into testing and training sets, with activity done as the output label – the training set is used to train the networks, and the testing set gives an accuracy of how well the network performs. If a model is predicting at a high accuracy (such as 80%), that model can be used for reliably.

*C. Machine Learning Models*

Five models were chosen for predicting activity – K Nearest Neighbor, Random Forest, Support Vector Machine, Artificial Neural Network, and Naive Bayes. Classifiers chosen followed a previous exercise classification study [11]. The networks were trained with various combinations of sensor data, and accuracies are to be gauged and compared.

*1) K Nearest Neighbor*

The K Nearest Neighbor (KNN) algorithm is the simplest classification algorithm of the list. It locates the *k* nearest neighbors to each input vector that share a label. Using the distance metric of (1), the Euclidean distance, the algorithm can compute proximities between all *n* dimensional points *p* and *q*, with *n* defined in the feature selection stage of the training area.

$$d(p,q) = \sqrt{\sum_{i=1}^{n}(q_i - p_i)^2} \quad (1)$$

The output of KNN is a class membership determined by majority vote of *k* closest neighbors, with $k \geq 1$. When $k = 1$, the closest neighbor to the input vector is the assigned label.

*2) Random Decision Forest*

Random Decision Forests (RDF) are constructed as an *ensemble* of decision trees. Each tree is independently trained from different subsets of the same training data. By averaging lots of trees the chance of overfitting the data greatly decreases. Training a random decision forest involves maximizing *information gain* at each node [12], which is defined as Entropy(parent) – Weighted Sum of Entropy(Children), with equation (2) formally defining entropy for tree node *n*, with *P(i)* as the fraction of classes with class *i* at node *n*.

$$H_n = -\sum_i P(i) \log_2 P(i) \quad (2)$$

*3) Support Vector Machine*

Support Vector Machines (SVM) classify data by calculating the optimal hyperplane separating two classes. When a new data point is received, the SVM predicts class ownership by determining what side of the hyper-plane the point falls on. By maximizing the margin between the hyperplane and closest training points, (also known as the support vectors), an optimal linear separation can be achieved. The general equation for a hyper-plane is given in (3), where *w* and *b* are the hyper-plane normal vector and offset, respectively. All points on the hyper-plane $x_p$ satisfy (4).

$$f(x) = w_1 x_1 + w_2 x_2 + \cdots + w_n x_n + b = 0 \quad (3)$$
$$f(x_p) = w^T x_p + b = 0 \quad (4)$$

*4) Naive Bayes*

This classification model uses conditional probabilities $p(C_k|x)$ of each class $C_k$, given input features *x*, with the most probable class returned as the prediction. Naive Bayes (NB) makes a strong independence assumption between features, resulting in a much simpler but less realistic model.

$$p(C_k|x) = \frac{p(C_k)p(x|C_k)}{p(x)} = \frac{p(C_k)}{p(x)} \prod_{i=1}^{n} p(x_i|C_k) \quad (5)$$

$$p(x_i|C_k) = \frac{1}{\sqrt{2\pi\sigma_k^2}} e^{-\frac{(x_i - \mu_k)^2}{2\sigma_k^2}} \quad (6)$$

The denominator of (5) is effectively a constant for a given feature vector and does not affect the final classification. The prior probabilities, $p(C_k)$ are calculated by looking at the occurrence probabilities of each class in the training set. The conditional probabilities, $p(C_k|x)$, are modeled using a Gaussian distribution over all feature/class pairs as shown in (6). The individual distributions are calculated by taking training data for each class/feature pair and calculating the mean and variance. These can then be used along with (6) to calculate conditional probabilities.

*5) Artificial Neural Networks*

The Artificial Neural Network [13] is a family of classifiers characterized by sets of nodes densely interconnected by weighted edges, capable of modeling non-linear, high-dimensional functions via a supervised learning algorithm. This study uses a neural network of two fully connected hidden layers with sigmoid activation function and a third output layer, with sigmoid function (7) at each output neuron.

$$\varphi(z) = \frac{1}{1 + e^{-z}} \quad (7)$$

Eq. (7) outputs [0, 1] - the largest value correlates to the predicted label.

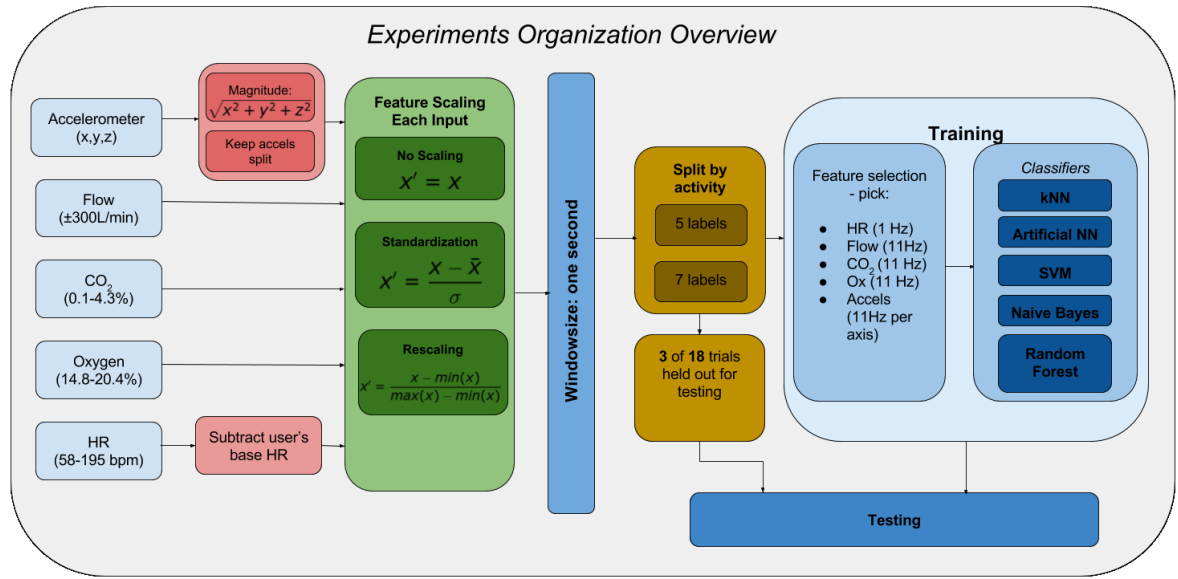

Fig. 4. Experiments organization, from raw sensor data to training and testing stage

## V. Experimental Evaluation

### A. Setup and Methodology

For this study eighteen trial members were instructed to carry out a fifteen minute set of exercises (Figure 2). Once the heart rate device is attached and the mask is turned on (refer to III.B), the trial can begin. The Android application is programmed to notify the user of transitions automatically, and stops at exactly the 900 second mark.

For the first two minutes (*maskoff*) of the trial the mask is laying on a flat surface, off the face, allowing the $CO_2$ and $O_2$ sensors to calibrate to atmosphere levels. All sensors are collecting data starting at the zero minute mark, and read close to zeros for flow and acceleration. Here the individual is standing still, and baseline heart rate is picked up.

At the two minute mark the mask is put on the face, continuing to stand. The next two minutes are walking, then two minutes of jogging, two minutes of running and one minute of jumping in place. After jumps there is two minutes of walking again (labeled as *walk2*) and standing (*stand2*). During the trial the administrator moves with the trial member, making sure the mask is properly positioned and the app collecting data. At the fifteen minute mark the app stops the data collection and all data is saved to a cloud data service (Firebase). The heart rate stopwatch is turned off as well.

### B. Data Pre-Processing

After all 18 trials have been performed, the raw data from the sensors is pulled from the cloud storage service, Firebase. Then heart rate data is exported from the Polar watch and extrapolated to fit the sampling of the sensors. A randomly selected profile is plotted in Figure 3.

Since only the period when the mask is on the face is of interest, the first two minutes of each trial are removed from the statistics calculations, leaving 13 minutes of activity for the training models.

The experiment organization follows Figure 4. To standardize heart rate values for each user, the first two minutes of each trial's heart rates are averaged and taken as their baseline heart rate and subtracted from the input heart rate vector.

Accelerometer data is taken from the three axes and optionally combined to get the magnitude of the acceleration. This vector provides a good idea of total head movement during activity, without keeping information of direction.

Optional data scaling follows. Standardization of the data is performed by applying (8) to each element of the vector.

$$x' = \frac{x - \bar{x}}{\sigma} \quad (8)$$

This transforms each input vector into a zero mean, unit variance vector, giving a vector of z-scores. The other type of standardization is feature scaling:

$$x' = \frac{x - \min(x)}{\max(x) - \min(x)} \quad (9)$$

Eq. (9) scales all data to the range [0, 1]. Each scaling method exactly preserves the shape of the original data graph, but provides useful information to the models when comparing among trials.

A window size needed to be chosen to set the input vector. Window sizes from 1 to 5 seconds in length were experimented with, to see if generalized error was increased, and it was noticed that a one second window gave performance on the same if not better level as larger sizes. A one second window provides an input vector from minimally length 1 (just one heart rate sample) to maximally length 67 (one heart rate sample along with 6 sensor samples at 11Hz).

Activity labels are either selected from a range of [1,7] or a range of [1,5]. As mentioned earlier in the section, there are 7 phases of activities in the 13 minutes of a trial. If we distinguish between the different phases of walking and standing, we have 7 labels for the activities. If we consider all walking and standing phases to be equivalent, we label the data with only 5 labels. Having these two sets of labels provides another opportunity to see if the models can distinguish between the same physical but somewhat different breath profiles.

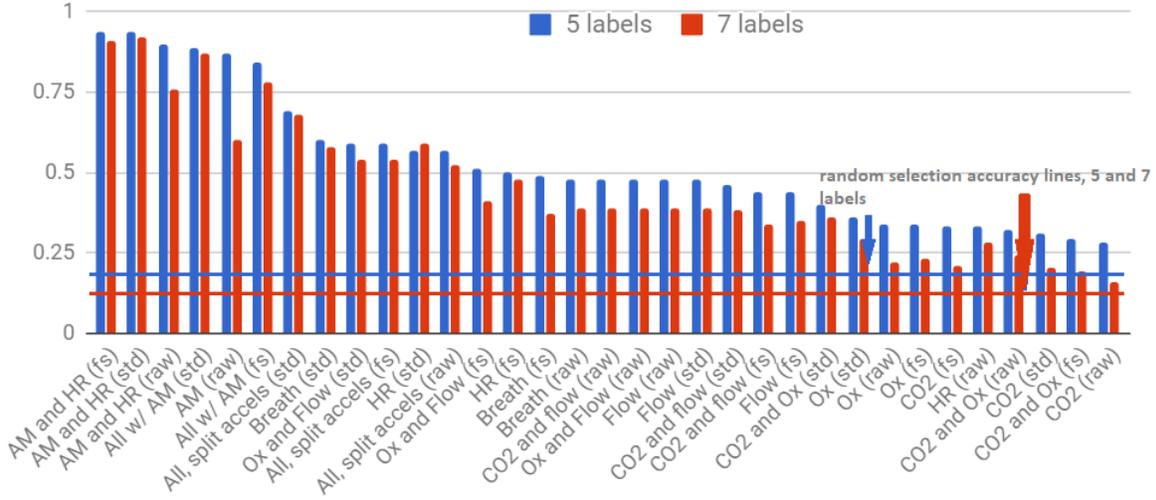

Fig. 5. Testing accuracies for various feature selections using KNN with k=10. Abbreviations: AM = acceleration magnitude, fs = feature scaling, std = standardization. Drawn horizontal lines represent accuracy using random selection.

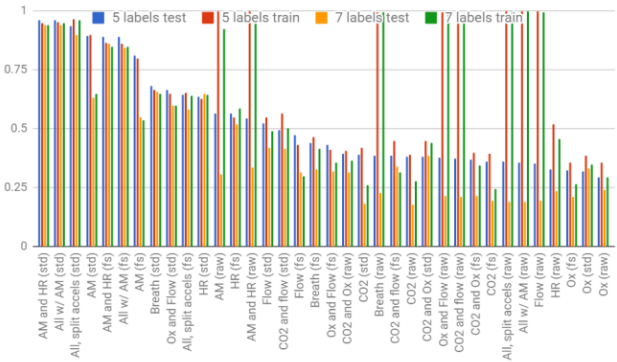

Fig. 6. Test & train accuracies for support vector machine (SVM).

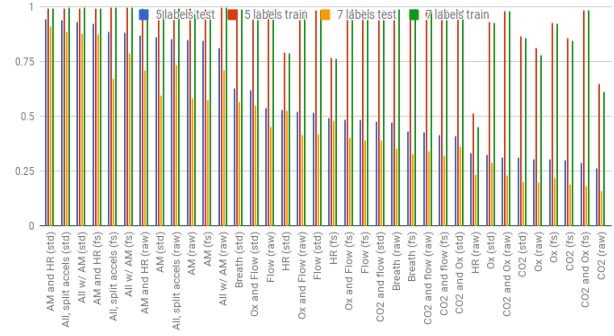

Fig. 7. Test & train accuracies for random decision forest (RDF).

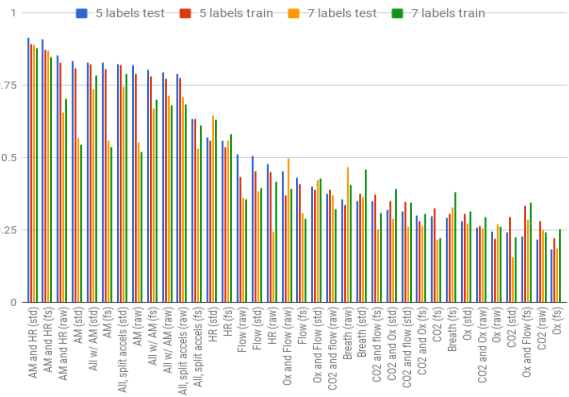

Fig. 8. Test & train accuracies for Naive Bayes (NB).

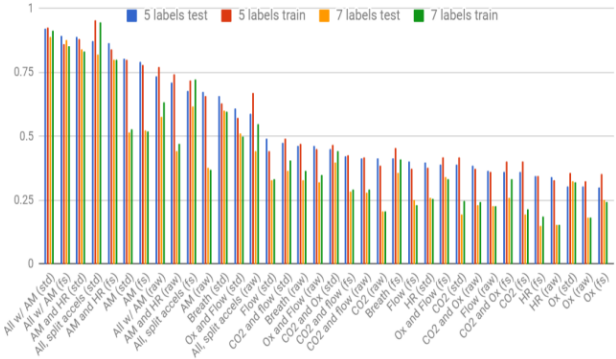

Fig. 9. Test & train accuracies for artificial neural net (ANN).

Because a person's transition from one phase to another is not instantaneous, five seconds of data around each transition are discarded from model training. For an activity of two-minute duration, 110 seconds of data are kept for training, (jogging retains 50 seconds).

### C. Results

Due to space constraints, the top two performing models are discussed in depth; however, all results are shown in Figures 5-9, with a summary in Figure 10.

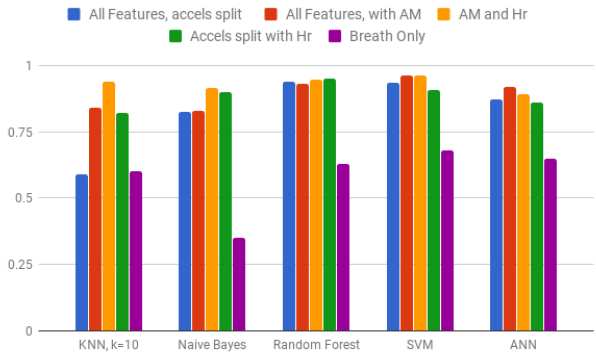

Fig. 10. Summary of testing accuracies for all models, using standardization.

SVM proved to be the best performing model - results of the training and testing phase of the SVM are shown in Figure 6. Optimal performance on 5 labels occurs when standardizing acceleration magnitude and heart rate, at 96.2%. SVM also provides the best breath-only prediction of all the models, with an accuracy of 68% and 65% for 5 and 7 labels, respectively.

Figure 7 summarizes RDF results. The RDF gives 94.5% accuracy for 5 labels using standardized acceleration magnitude and heart rate, the second-best performance. Standardized breathing gives 63% and 57% for 5 and 7 labels, respectively. Standardization typically performs better than feature scaling, which performs better than raw data. Also of note is the fact that the training accuracy scores are near perfect – a feature of random decision forests.

*D. Discussion*

The best-scaled results are summarized in Figure 10 for 5 labels. Using all features with acceleration magnitude, the best testing accuracy comes from SVM at 96.2%. Using just breathing features, the accuracy is 68% and 65% for 5 and 7 labels, respectively.

The best prediction results come from using acceleration (either magnitude or split) along with heart rate. This combination gives accuracies of >85% for all the models (usually >90%). Using acceleration to predict physical activities was expected to yield high accuracies – previous studies have used standalone and smartphone accelerometers to predict activity with >90% accuracy [14, 15].

Breath only data (flow, $CO_2$, and $O_2$) on 7 activities gives a 65% prediction result. This accuracy result is promising, but not quite high enough to be used with great confidence. Also of note is the fact that flow data is the more dominant predictor of activity compared to the gas percentages.

We believe that higher accuracy can be achieved using breath-only statistics, but will require more highly calibrated mask sensors with a carefully designed, air tight mask. Also, having a growing patient data set delimited by age, weight, gender, and health condition would add more parameters to training the network and would likely improve accuracy. In part this is because maximum oxygen uptake, for example, is linearly proportional to age and varies with gender [16].

## VI. Conclusion

The purpose of this study is to see how accurately machine learning models can predict user activity based on breathing data, and if they can achieve the same accuracies as acceleration and heart rate. From this study one can see that several machine learning algorithms, especially SVM and Random Decision Forest, can be used to predict activity using breathing data with an accuracy of >65%. We hope that this analysis will lead future research into activity prediction using breath, with potentially more breakdown by age, gender, and weight. Such research could be helpful for troubleshooting a patient's breathing data without complicated, in-house treatments. With time such networks could help doctors pinpoint lung cancer patients susceptible to problematic resectable tumor operations.

References

[1] "Key Statistics for Lung Cancer", *Cancer.org,* 2017. [Online]. Available: https://www.cancer.org/cancer/non-small-cell-lung-cancer/about/key-statistics.html.

[2] Kearney DJ, Lee TH, Reilly JJ, DeCamp MM, Sugarbaker DJ. Assessment of operative risk in patients undergoing lung resection. Importance of predicted pulmonary function. Chest [Internet]. 1994;105(3):753–9. Available from: http://journal.publications.chestnet.org/data/Journals/CHEST/21691/753.pdf

[3] Colice GL, Shafazand S, Griffin JP, Keenan R, Bolliger CT. "Physiologic Evaluation of the Patient With Lung Cancer Being Considered for Resectional Surgery". Chest [Internet]. The American College of Chest Physicians; 2007;132(3):161S – 177S. Available from: http://dx.doi.org/10.1378/chest.07-1359

[4] "Cardiopulmonary Exercise Test Procedures", *Stanfordhealthcare.org,* 2017. [Online]. Available: https://stanfordhealthcare.org/medical-tests/c/cardiopulmonary-exercise-test/procedures.html.

[5] Jones, Campbell. Clinical Exercise Testing, 2nd edition. W. B. Saunders Company, 1975.

[6] B. Myers, J. Nahal, C. Yang, L. Brown, S. Ghiasi and A. Knoesen, "Towards Data-Driven Pre-Operative Evaluation of Lung Cancer Patients: The Case of Smart Mask", IEEE Wireless Health Conference, 2016.

[7] H. Ghasemzadeh, "Structural Action Recognition in Body Sensor Networks: Distributed Classification Based on String Matching", IEEE Transactions on Information Technology in Biomedicine, Vol. 14, 2010

[8] A. Yang, R. Jafari, S. Sastry, R. Bajcsy, "Distributed Recognition of Human Actions using Wearable Motion Sensors", Journal of Ambient Intelligence and Smart Environments, Vol 1, no. 2, pp 103-115, 2009

[9] H. Ghasemzadeh, R. Jafari, "Physical Movement Using Body Sensor Networks: A Phonological Approach to Construct Spatial Decision Trees", IEEE Transactions on Industrial Informatics, 2011

[10] R. Jafari, A. Encarnacao, A. Zahoory, F. Dabiri, H. Noshadi, M. Sarrafzadeh, "Wireless Sensor Networks for Health Monitoring", Mobile and Ubiquitous Systems: Networking and Services, 2005.

[11] N. Hosein and S. Ghiasi, "Wearable Sensor Selection, Motion Representation and their Effect on Exercise Classification", *IEEE First Conference on Connected Health*. IEEE, 2016, pp 370-379.

[12] Witten, Ian; Frank, Eibe; Hall, Mark (2011). Data Mining. Burlington, MA: Morgan Kaufmann. pp. 102–103.

[13] S. Haykin, Neural Networks and Learning Machines, 3rd edition. Pearson, 2008.

[14] J. Kwapisz, G. Weiss, S. Moore, 2010. "Activity Recognition Using Cell Phone Accelerometers", *IEEE Biometrics Compendium*

[15] N. Ravi, N. Dandekar, P. Mysore, M. Littman, 2005, "Activity Recognition from Accelerometer Data", *17th Conference on Innovative Applications of Artificial Intelligence, Vol 3*, pp 1541-1546.

[16] V. Heyward. Adv. Fitness Asst & Exercise Prescription, 3rd ed 1997.